# RFID EXPLOITATION AND COUNTERMEASURES

L. Gavoni, *EC-Council University*

*Abstract*- **Radio Frequency Identification (RFID) systems are among the most widespread computing technologies with technical potential and profitable opportunities in numerous applications worldwide. Further, RFID is the core technology behind the Internet of Things (IoT), which can accomplish the real-time transmission of information between objects without manual operation. However, RFID security has been taken for granted for several years, causing multiple vulnerabilities that can even damage human functionalities. The latest ISO/IEC 18000-63:2015 standard concerning RFID dates to 2015, and much freedom has been given to manufacturers responsible for making their devices secure. The lack of a substantial standard for devices that implement RFID technology creates many vulnerabilities that expose end-users to elevated risk. Hence, this paper gives the reader a clear overview of the technology, and it analyzes 23 well-known RFID attacks such as Reverse Engineering, Buffer Overflow, Eavesdropping, and Malware. Moreover, given the exceptional capabilities and utilities of RFID devices, this paper has focused on security measures and defenses for protecting them, such as Active Jamming, Shielding tag, and Authentication.**

I. INTRODUCTION

The origins of the chips date back to the dawn of the Second World War where the British, unlike many other nations, began to appreciate the benefits obtained from the prototypes of the radar in the maritime field. Therefore, it was decided to develop a more modern and effective radar system to anticipate a possible war against Germany. This radar had to recognize the enemies even when they were not yet in sight. This was essential, especially in the naval field (since the fleet and the Royal Air Force were the only effective defense for England in case of invasion). It was decided to develop *computerized* systems capable of doing this. No one expected an attack before three years, so the experiments started at the beginning of 1939, proceeded with relative tranquility until Germany invaded Poland on September 1 of the same year. Despite this (thanks to the policy of appeasement wanted by Prime Minister Neville Chamberlain), the experiments proceeded slowly until the beginning of the Battle of Britain (spring-summer 1940). Here Prime Minister Winston Churchill gave a boost to the scientists/engineers who sternly started to work. The first chips capable of reaching the goal sought were not available until the end of 1944 and were installed as a test in some fighter planes that took part in the last offensives of World War II, as shown in Fig. 1.

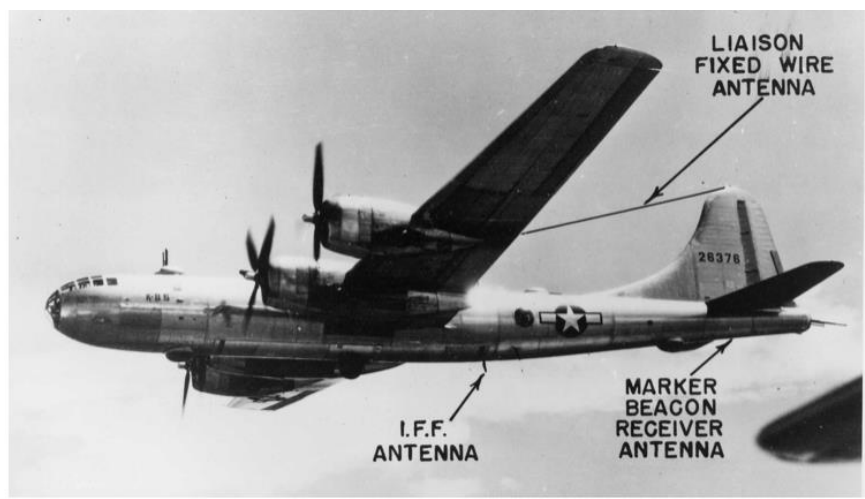

Figure 1. Portside antennas on Boeing yb-40 flying fortress [1]

This happened when the Germans unleashed Operation Wacht Am Rhein (the Ardennes Offensive or Battle of the Bulge) during the winter of 1944-45, an extreme attempt to drive the Allies back into the sea using everything they had at their disposal. In this case, the chips proved helpful, especially in reduced visibility, although not always reliable. However, the first tests were satisfactory, which led to further development of the technology that proved to be very useful since the Korean War.

*A. Technology fundamentals*

RFID stands for Radio Frequency Identification, an identification technique based on radiofrequency. It is a technology that allows the unique identification of an object. Indeed, it transmits the identity of an object in the form of a unique serial code using radio waves.

The basic concept underpinning the technology is straightforward:

1) We place a transponder, a microchip with an antenna, inside an object, for example, a sheet.
2) We use a Reader, a device with one or more antennas, to read the data entered the microchip using radio waves. To succeed in the intent, the reader and the microchip will be synchronized to a particular standard frequency. The reader will emit radio waves, and thanks to the physical phenomenon of *magnetic induction*, the chip will be "activated" by the antenna's magnetic field. The tag[1] can accumulate or amplify the energy it needs to transmit the information it contains.
3) Finally, after the reader collects the data from the tag, it transfers the data to the RFID controller by a network connection (Ethernet, Mobile IP, etc.). The controller will take the information and use it depending on the type of data it has received. The data can be collected in a database or moved as an object in the inventory.

The above procedure can be represented in Fig. 2 to make the concept clearer.

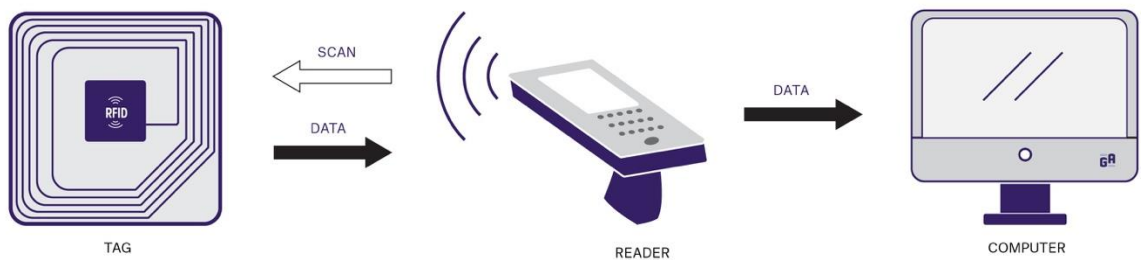

Figure 2. RFID System Architecture [2]

The surprising peculiarity of RFID tags is given by the extraordinary ability to combine tiny dimensions with massive memory capacity.

---

[1] Electronic device that holds data. Chip-based RFID tags contain silicon chips and antennas.

*B. RFID Classification*

The communication frequencies between the reader and the tag depend on the structure of the tag and the intended applications and are regulated by the usual international and national bodies. Regulation, however, is divided into geographic regions with different laws from region to region, which often results in incompatibility when RFIDs travel to different nations along with the goods with which they are associated. The portions of frequency bands most used in RFID technology are:

- LF (Low Frequencies) band: in particular, the sub-band 120-145 kHz. It is located in the lowest part of the RF spectrum and is historically the first frequency used for automatic identification, and still has a significant presence in the market. It is used for access control, animal identification, and inventory control.
- HF (High Frequencies) band: sub-band centered on 13,56 MHz. It is considered the "universal" frequency band, usable worldwide, making it the most widespread band to date. It is used for access control and smart cards.
- UHF (Ultra High Frequencies) band: it is the *new band* for RFID, with a much more comprehensive range of operation than LF and HF. Unfortunately, the band is not uniformly assigned in the various countries, and it is included between 850-950 Mhz.
- SHF (Super High Frequency) band: it is used by systems that operate using Microwave bands in frequency ranges around 2.45 GHz and 5.8 GHz. It is based on far-field radiative coupling or backscatter coupling principles. It has short wavelengths and can thus be used with metals.

Further, the RFID tags can be classified based on the use of a power supply:

- Passive: Passive tags use the field generated by the reader's signal as a power source to power their circuits and transmit. However, the power that can be derived from the reader's signal is not only low, but it decreases very rapidly with distance and is limited by the reader's RF emission level regulations. This results in relatively low operating distances (a few meters at most). On the other hand, they are preferred when tiny transponder dimensions are needed. The main technological goals are low consumption and the ability to handle RF signals affected by noise. Tags contain a certain amount of non-volatile memory EEPROM, and for a universal identifier (EPC), at least 96 bit are required. The more memory, the larger the chip size and the higher the cost. This type of tags is the most widespread and used in massive applications and is realized on almost all frequency bands allowed for RFID applications. Examples of passive tags are the chips used in smart cards and the laundry of industrial laundries.
- Active: Active tags have their power supply, typically a battery and a radio frequency transmitter/receiver. Usually, the memory onboard is more significant than passive tags, and reading and writing operations can be performed on it. Another advantage of active tags is the much longer operating distance than passive ones, as they are equipped with an accurate transmitter powered by an energy source. The reachable distance is limited only by the antenna and the energy available in the batteries. It can reach a radius of a few kilometers. The cost of these devices can reach tens of dollars, and they generally use the ultra-high frequencies (UHF) band. Further, they are naturally dedicated to *valuable* applications, or in cases where the tag is reusable several times. Examples of active chips are the unique bracelets used in hospitals to control patients with specific diseases (such as Parkinson's and Alzheimer's).
- Semi-passive: In between active and passive RFID tags are semi-passive tags or battery-assisted passive RFID tags. They look more like passive tags in terms of size and ease of manufacture. Still, like active tags, they incorporate a power source (usually a tiny, eco-friendly battery) to improve data transmission. The result is a tag that offers many of the benefits of active RFID tags, but at a price point that is closer to a passive tag.

In the next section, we will focus on the types of attacks aimed at damaging the privacy and security of RFID technology.

## II. RFID ATTACKS

RFID technology is challenged by numerous security and privacy threats that make the technology very simple to exploit if the correct security measures are not implemented. "*Since the RFID reader communicates with electronic tag through wireless channels, and the security of wireless channel transmission is relatively weak, the information transmitted between the reader and the tag is vulnerable to eavesdropping, tampering, and other attacks*" [3].

RFID security attacks can be classified into two main categories: privacy violations and security violations. In privacy violations, the attacker seeks to harvest information from the objects by eavesdropping on the communications between the object and the reader or by tracking them. On the contrary, in security violations, a cybercriminal counterfeits the behavior of a tag or a reader for making undesirable communications. To have a clear picture of the RFID attacks, the reader can see in Fig. 3 a graphic representation of the various types of attack based on the level of the affected surface, which will be explained in the following paragraphs. Multilayer attacks also exist, which affect more than one layer.

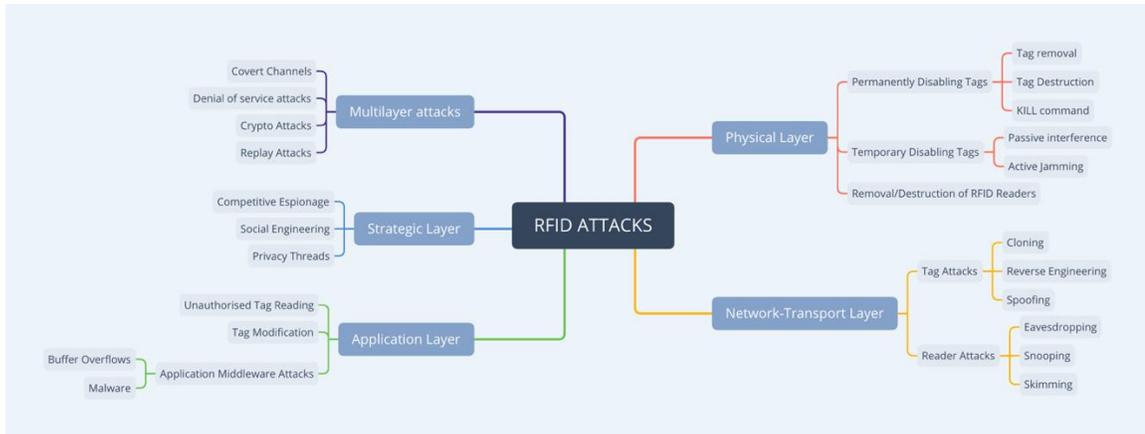

Figure 3. RFID attacks overview

### C. Physical layer

Physical threats use physical means to attack the RFID system to disable tags, modify their content, or emulate them. Indeed, since RFID tags present poor physical security, RFID tags that are not embedded in items can easily be removed from an item and may subsequently be attached to another one. For instance, remove the tag from an item with a high price and switch it to another item with a low price. Also, a thief could damage an RFID tag and afterward take the product carrying it through an automated checkout portal without the store detecting that the product has been taken out. Moreover, RFID tags are sensitive to static electricity since their electronic circuits can be damaged in a brief time by electrostatic discharge caused by conveyor belts or high energy waves. Also, the Auto-ID center and EPC global created a command specification called KILL that can permanently silence an RFID tag. According to this system, each RFID tag has a unique key that is defined by the tag's manufacturer, and its use can make an RFID tag permanently inoperable. Depending on the type of deactivation used, the KILL command may also partially or completely erase any data stored on the device. Furthermore, it is also possible to disable the tag in a non-permanent way. An attacker can use a Faraday cage such as wire mesh or metal plates to block electromagnetic waves from it. In other cases, the attacker may prevent tags from communicating with readers by generating a signal in the same range as the reader, called active jamming. Finally, the reader must consider that *RFID systems*' communication is rendered susceptible to possible interference and collisions from any source of radio interference such as noisy electronic generators and power switching supplies. This interference, called passive interference, prevents accurate and efficient communication between the tags and the readers [4].

D. *Network-Transport Layer*

This layer covers all the attacks based on how the RFID systems interact and whereby data are transferred between the entities of an RFID network. Hence, this section outlines attacks that affect the network transport layer. For the sake of clarity, they are divided into attacks on the tags and on the reader.

E. *Tag Attacks*

Since the cost of an RFID tag is usually cheap, it is not challenging to make cloning. An attacker with physical access to a tag can duplicate a tag with reverse engineering. Each RFID tag uses for the identification a unique hardcoded ID number. If the attacker exposes the ID information, the tag can easily be copied. Further, the degree of effort needed to accomplish this attack depends on the security features of the RFID tags. Thereafter the RFID system will be accessed using this identity information to impersonate the legitimate identification number or readers, which is defined as counterfeiting or spoofing attacks.

F. *Reader Attacks*

"*The wireless nature of RFID makes eavesdropping one of the most serious and widely deployed threats*" [5]. Eavesdropping occurs when the channel is secretly listened to by an attacker to retrieve information from it. Since UHF RFID systems cover more reading distance than other frequency bands, this attack is more likely to happen in them. Eavesdropping is a viable threat and hard to be identified since it can be carried out at a more extended range on the communications between a tag and a valid reader. At the same time, the adversary is passive and does not send out any signal. A similar attack is called snooping, which occurs when the data stored on the RFID tag is read without the owner's awareness or agreement by an illegal reader interacting with the tag. This attack happens because almost all the tags transmit their stored data in their memory without authentication. Lastly, if the attackers study the observed data from the reader to the tags, they can make a cloned tag that imitates the original RFID tag. This attack is called skimming and is very popular where any license or passport is scanned through an RFID system.

G. *Application Layer*

The Application layer comprises all the attacks that target applications and the binding between the final users and RFID tags. Hereafter will be described the unauthorized tag reading, tag modification, and attacks in the application middleware.

Since there is no on/off switch for RFID devices and not every tag support authenticated read operation, an attacker may easily read the contents of RFID tags without leaving any trace, thus achieving an unauthorized tag reading attack. Also, an attacker can perform a tag modification attack. Indeed, since most RFID tags use writable memory, an adversary can benefit from this vulnerability to modify or delete relevant data from the tag's memory.

This attack's impact will lean on the application in which the tags are utilized and the degree to which tag data are modified. Consequently, the discrepancy among data stored on the RFID tag and the corresponding tagged object may have profound implications; suffice it to think of hospitals.

## H. Middleware attacks

A buffer is a sequential section of memory that contains anything from a character string to an array of integers. A buffer overflow, or buffer overrun, occurs when more data is put into a fixed-length buffer than the buffer can handle. The extra information, which must go somewhere, can overflow into adjacent memory space, corrupting or overwriting the data held in that space [6].

Hence, the attackers may use RFID tags to launch buffer overflows on the back-end RFID middleware if the system is poorly designed. Thus, they may insert code or commands in memory buffers read by processes that can execute administrative functions disabling security controls.

Malware represents another middleware attack. Malware, or malicious software, means several malicious software variants, including viruses, ransomware, and spyware. A malware commonly consists of code developed by cybercriminals designed to cause widespread damage to data and systems or gain unauthorized access to a network. In the past, it was not thought feasible to think of malware within RFID devices, given the limited memory available within them. Nevertheless, in research from [7], the first RFID-related malware has been demonstrated. In Fig. 4, it is possible to denote the basic structure of an RFID malware thread model.

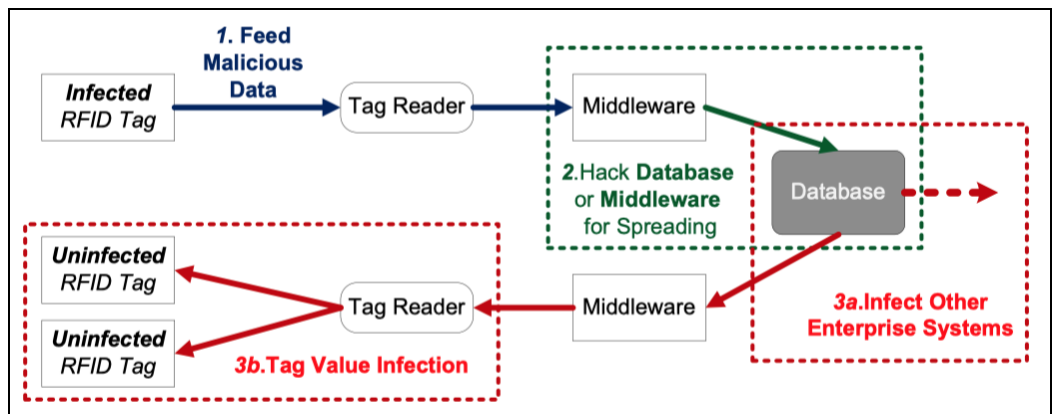

Figure 4. Essential Thread Model for RFID Malware [7]

The infected RFID tags can be common tags with limited resources or more powerful tags such as contactless smart cards. First, RFID tags infected with malicious data are given as input to the tag reader. The data is used to exploit the vulnerability of the database system or the RFID middleware. Finally, if the attack is successful, the malware can be spread by refreshing tag values with infected data during regular tag updating.

## I. Strategic Layer

In this layer, it is possible to find several attacks that target organizations and business applications, taking advantage of the negligent design of infrastructures and applications.

An attack is often sought and paid for by a competitor company to obtain sensitive and confidential data such as availability stocks and marketing secrets. This type of attack is called competitive espionage. Such attacks can be accomplished via eavesdropping or by gaining unauthorized access to back-end databases. In addition, an attacker can manipulate some people physiologically to extrapolate confidential information. The degree of success of the attack is mainly based on the attacker's social engineering skills. Also, since RFID tags react to any reader, trustworthy or not, this feature can be exploited by attackers to trace and profile individuals. This creates a privacy thread, and the severity of the attack depends solely on the owner's information, such as sensitive data.

*J. Multilayer attacks*

Lastly, in this category are covered attacks that target multiple layers, including the physical, the network-transport, the application, and the strategic layer. Imagine that few people are admitted for surgery and paired with an RFID bracelet with their ID on it. What if their bracelet contains information that they do not know, such as medical or personal data? This type of attack is called covert channel and can be exploited by an attacker using an unauthorized communication channel to transfer secretive information.

RFID systems may also be subject to denial of service (DoS) attacks, which lets the system not work correctly. The attacker targets to block the reader from reading tags by using a blocker tag or the LOCK commands. Over time, several solutions were made to defeat the DoS attacks in all the other communication systems. Nevertheless, they cannot be used in RFID systems due to the limited resources of RFID tags. Furthermore, since RFID middleware includes networking devices, a cybercriminal may take advantage of the system's faulty resources and perform a denial of service in the RFID middleware. Typically, different encryption algorithms are used to secure the data if the RFID chips contain sensitive data (such as a passport). Still, an attacker can perform a crypto attack to break the algorithm and reveal sensitive information.

Most attacks shown before were based on functional vulnerabilities. Now will be described in detail the side channel attack that uses a lack in the physical implementation of a cryptographic algorithm. Specifically, in a side-channel attack, a cybercriminal relies on the information harvested from the physical implementation of the cipher, such as timing analysis, power monitoring, fault attack, and electromagnetic radiation (Differential Power Analysis). One of the most effective attacks is power analysis since the power consumption of a cipher may provide much information about the running operations and their involved parameters. Further, this attack needs effortless equipment such as a PC with an oscilloscope and a small resistor in the power supply line to measure the power.

Finally, the last attack we will present is called replay attack. A cybercriminal collects by eavesdropping several response messages from the RFID communication channel and then reuses them for illicit purposes such as entering a building.

In the next section, we will focus on the security mechanisms and defense to protect the privacy and security of RFID technology.

## RFID Security Measures and Defences

To address the numerous aforementioned security threats, RFID devices had to employ various security measures designed to counter the different threats. In the following, we will describe several mechanisms that have been proposed to enhance the RFID's security and privacy. Effective mechanisms should provide protection against the risks mentioned. Nevertheless, the cost of the approaches should also be considered.

Over the years, several security implementations have been developed that can significantly reduce the risk of a cyber-attack. Hereafter we may find them:

- Sleeping and killing tags: the RFID tags are "killed" upon purchase of the tagged product by a buyer. Once the tag is killed by sending a particular command, including a short password, it is no longer functional and cannot be re-activated anymore. For instance, in a clothing shop, the tags of purchased goods would be killed at checkout to protect consumers' privacy. Therefore, none of the purchased items would contain alive RFID tags. However, since the tag cannot be reused in this method, its lifetime is limited and cannot be utilized for after-sale purposes. Sleeping tag is an improved approach similar to killing tag, but its merit is that the deactivated tag can be activated by *wake*.

- Shielding Tag: Faraday cage is a straightforward way of protecting an RFID tag inspired by the characteristics of electromagnetic fields. It isolates RFID tags from any electromagnetic waves. Indeed, since any external radio signals cannot penetrate the cage, no reader can access the tag to read it as long as the RFID tag is inside such a cage. A different approach to shielding tags is the active jamming approach, which isolates the tags from electromagnetic waves by disturbing the radio channel.
- Blocker Tags: a blocker tag is comparable to an RFID tag with the exception that it can block readers from reading the identification of those tags that exist in its range. It works by creating collision for a reader when it is attempting to identify tags in its field. This way, blocker tags can establish a safe zone around the tags, and all RFID tags that exist in this zone can impede reading their data in the presence of a blocker tag. With this solution, it is possible to protect consumer privacy with a relatively inexpensive investment. Still, an attacker cannot access tags just in a defined range, and beyond this range, tags are not shielded from attacks.
- Tag Pseudonyms: it uses a tiny pseudonyms collection and rotates these pseudonyms as its identifier in every tag, releases a different one on each reader query. The authorized readers share the full pseudonym set with the tag in advance to identify the tag. Since attackers cannot associate two different pseudonyms of the same tag, it would be more challenging for unauthorized tags to track.
- Proxy Privacy Devices: typically, RFID readers and tags cannot have the ability to provide consumer privacy protection. One way to overcome this challenge is to rely on the reader for privacy protection. Nevertheless, relying on the reader for privacy is risky since the reader is public. However, users may carry their own privacy-enforcing devices for RFID instead of relying on public RFID readers to enforce privacy protection. Indeed, creating a platform that offers centralized RFID security and privacy management for individual people.
- Authentication: it is a process through which an object demonstrates its claimed identity to another communication party, providing some evidence such as what it knows, what it has, or what it is. [9] stated a basic PRF private authentication scheme for mutual authentication between tags and readers. This protocol uses a shared secret and a Pseudo-Random Function (PRF) to protect the messages exchanged between the tag and the reader. Also, they proposed a tree-based private authentication and delegation tree scheme to reduce the server's load in the hash schemes.

Besides the mechanisms outlined above, the National Institute of Standards and Technology of the U.S. (NIST) designed the guidelines for securing RFID systems. These security controls are classified into three groups: management, operational, and technical, which describe the considerations and controls in detail [8].

## Summary

RFID can be a valuable tool in everyone's life if adequately implemented, as shown in this paper. Starting from a brief introduction of RFID history, the RFID classifications and architecture were defined and explained in detail, thus presenting possible technology fields that could currently be implemented. Since there are outdated RFID standards, manufacturers have too much freedom of choice, which often results in a lowering of the security of their devices. Hence, 23 well-known RFID attacks were described in detail, leaving the reader a complete overview of the potential vulnerabilities. In the last chapter, several security measures and defenses were introduced objectively, leaving the reader with reasoning ideas for the proper mitigation to apply according to their own needs.

To conclude, RFID devices have a significant impact in multiple industries, such as hospitals and retail. The safety of their customers must be put first, trying to increase the security of the devices and invest in researching new technologies that are increasingly secure. Furthermore, there will be many implementations that the reader could study and analyze in the coming years.

## REFERENCES

[1] Aafradio. (2018). Port side antennas [Photograph]. Port Side Antennas. https://aafradio.org/NASM/B-29_antennas_port.JPG.
[2] Labtag. (2018). https://blog.labtag.com/everything-you-need-to-know-about-rfid-technology/ [RFID Technology Presentation]. RFID Technology Presentation. https://blog.labtag.com/wp-content/uploads/2020/03/RFID20Technology20-20Presentation20-207202072820x2025020px20.jpg
[3] Wang, Y., Shen, J., Guo, X., & Dong, W. (2020). Research on RFID attack methods. 2020 IEEE 3rd International rence on Automation, Electronics and Electrical Engineering (AUTEEE), 433–437. https://doi.org/10.1109/AUTEEE50969.2020.9315712
[4] Khattab, A., Jeddi, Z., Amini, E., & Bayoumi, M. (2017). Rfid security. Springer International Publishing. https://doi.org/10.1007/978-3-319-47545-5.
[5] Mitrokotsa, A., Rieback, M. R., & Tanenbaum, A. S. (2009). Classifying rfid attacks and defenses. Information Systems Frontiers, 12(5), 491–505. https://doi.org/10.1007/s10796-009-9210-z
[6] Veracode. (2020). What Is a Buffer Overflow? Learn About Buffer Overrun Vulnerabilities, Exploits & Attacks. https://www.veracode.com/security/buffer-overflow.
[7] Rieback, M., Crispo, B., & Tanenbaum, A. (2006). Is Your Cat Infected with a Computer Virus? Fourth Annual IEEE International Conference on Pervasive Computing and Communications (PERCOM'06). https://doi.org/10.1109/percom.2006.32
[8] Karygiannis, A., Eydt, B., Barber, G., Bunn, L., & Phillips, T. (2007). Guidelines for securing Radio Frequency Identification (RFID) systems. NIST. https://doi.org/10.6028/NIST.SP.800-98.
[9] Molnar, D., & Wagner, D. (2004). Privacy and security in library RFID. Proceedings of the 11th ACM conference on Computer and communications security - CCS '04, 210–219. https://doi.org/10.1145/1030083.1030112